\documentclass[twoside,12pt]{cmft}
\usepackage{amsfonts}
\usepackage{amssymb}
\usepackage{amscd}
\usepackage{amsmath}
\usepackage{epsfig}
\usepackage{graphicx, subfigure}
\usepackage{latexsym}

\newtheorem{remark}{Remark}
\newtheorem{lemma}{Lemma}

\begin{document}
\title{Efficient computation of the branching structure of an
algebraic curve}

\author{J.~Frauendiener }
\email{joergf@maths.otago.ac.nz}
\address{Department of Mathematics and Statistics, 
University of Otago,      
P.O. Box 56, Dunedin 9010, New Zealand\\
and: Centre of Mathematics for Applications, 
University of Oslo, 
P.O. Box 1053, Blindern, 
NO-0316 Oslo,
Norway}
\author{C.~Klein}
    \email{Christian.Klein@u-bourgogne.fr}
\address{Institut de Math\'ematiques de Bourgogne,
		Universit\'e de Bourgogne, 9 avenue Alain Savary, 21078 Dijon
		Cedex, France}
\author{V.~Shramchenko}
\email{Vasilisa.Shramchenko@Usherbrooke.ca} 
\address{D\'epartement de math\'ematiques, Universit\'e de Sherbrooke, 2500, boul. de l'Universit\'e 
Sherbrooke (Qu\'ebec) Canada J1K 2R1}
\date{\today}    
\thanks{This work has been supported in part by the project FroM-PDE funded by the European
Research Council through the Advanced Investigator Grant Scheme, the Conseil R\'egional de Bourgogne
via a FABER grant and the ANR via the program ANR-09-BLAN-0117-01. JF 
and VS thank for the hospitality at the IMB as visiting 
professors, where part of the work was written. The work of VS 
was supported by NSERC
}

\begin{abstract}
  An efficient algorithm for computing the branching structure of a
  compact Riemann surface defined via an algebraic curve is
  presented. Generators of the fundamental group of the base of the
  ramified covering punctured at the discriminant points of the curve
  are constructed via a minimal spanning tree of the discriminant
  points. This leads to paths of minimal length between the points,
  which is important for a later stage where these paths are used as
  integration contours to compute periods of the surface.  The
  branching structure of the surface is obtained by analytically
  continuing the roots of the equation defining the algebraic curve
  along the constructed generators of the fundamental group.
\end{abstract}

\keywords{Riemann surfaces, algebraic curves, monodromies, 
fundamental group}

\maketitle

\section{Introduction}
Riemann surfaces have many applications in natural sciences and
engineering, for instance in the solutions of certain integrable
partial differential equations (PDE) appearing in hydrodynamics and
optics, see e.g.~\cite{algebro}.  For a long time the full potential
of related techniques could not be realized due to the absence of
efficient numerical approaches. In \cite{deho}, the Maple package
\emph{algcurves} (starting with Maple 7) for algebraic curves which
gives a mixed symbolic-numeric approach was published, see also
\cite{depa}. Since all compact Riemann surfaces can be defined via
non-singular plane algebraic curves (see e.g.~\cite{springer}), all
quantities characterizing a Riemann surface can in principle be
computed along these lines. For a different numerical approach to
Riemann surfaces based on Schottky uniformizations see
\cite{bo,schmies}.

Though being very useful, the mixed symbolic-numeric approach has the
disadvantage that only algebraic curves with exact arithmetic
coefficients, i.e., not floating point coefficients can be used, and
that the performance of the numerics is consequently reduced in
addition to the limitations imposed by the restriction to exact
arithmetic expressions. Thus a fully numeric approach was presented in
\cite{cam,lmp} for real hyperelliptic curves and in \cite{frkl} for
general algebraic curves.  The gain in performance and in flexibility
allows the study of higher genus curves and of families of curves, i.e.,
of the modular properties of Riemann surfaces. Such modular
dependences are important for instance in the study of solutions to
certain integrable PDE appearing in the context of gravity and surface
theory \cite{ernstbook} and the description of the asymptotic behavior
of highly oscillatory regimes in dispersive PDE \cite{lale,cpam} as
well as the study of modular invariants discussed in topological field
theories, see for instance \cite{Sarnak,klkoko}.

A plane algebraic curve
$C$ is defined 
as a subset in $\mathbb{C}^{2}$, $C = \{(x,y)\in\mathbb{C}^{2}|f(x,y)=0\}$, where $f(x,y)$ 
is an irreducible polynomial in $x$ and $y$,
\begin{equation}
    f(x,y) = 
    \sum_{i=1}^{M}\sum_{j=1}^{N}a_{ij}x^{i}y^{j}=\sum_{j=1}^{N}a_{j}(x)y^{j} =0
    \label{fdef}\;.
\end{equation}
We assume that not all $a_{iN}$ vanish and that $N$ is thus 
the degree of the polynomial in $y$. At a generic point $x$ there are 
$N$ distinct roots $y^{(k)}$, $k=1,\ldots,N$, which implies that the 
algebraic curve defines an $N$-sheeted ramified covering of the 
$x$-plane. The surface is then compactified in a standard way (see 
for example \cite{BK}) so that we have a ramified covering of the $x$-sphere $\mathbb{CP}^{1}$. 
At a point where both $f(x,y)=0$ and $f_{y}(x,y)=0$, the number of 
distinct roots $y^{(k)}$ is lower than $N$, i.e., this
\emph{branch point} belongs to several sheets of the covering. To describe the associated  
Riemann surface, one has to be able to identify the branching 
structure of the curve at the branch points, in other words, one has to specify which sheets 
of the covering are connected in which way at a given branch point. 
This is equivalent to identifying the \emph{monodromy} of the 
surface. 
It is the purpose of this paper to give an efficient algorithm for 
this crucial step in the numerical treatment of Riemann surfaces. 
The structure of the Riemann surface obviously does not depend on 
whether the algebraic curve (\ref{fdef}) is studied  as a covering of 
the $x$- or of the $y$-sphere. We will concentrate here  on the covering 
of the $x$-sphere since this covering appears as the input data in many applications such as algebro-geometric solutions to certain 
integrable equations. Notice that the inverse problem, to find the 
equation of an algebraic curve for a given monodromy, is very involved 
and could so far be only addressed for low genus, see e.g.~\cite{couv} and 
references therein.

\begin{figure}
[!htbp]
\begin{center}
\includegraphics[width=6cm]{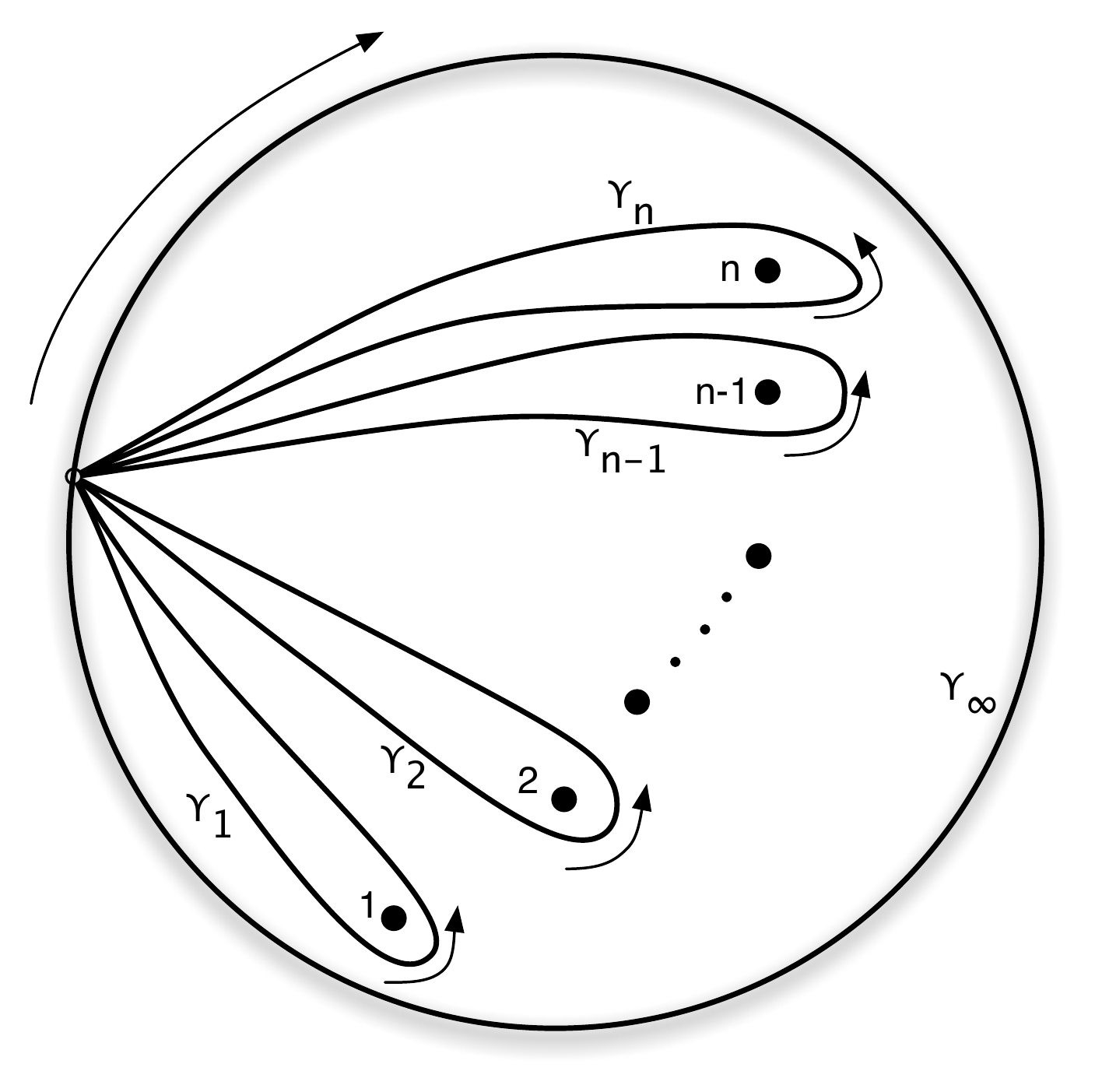} 
\caption{Generators of the fundamental group of 
$\mathbb{C}\setminus\{b_{1},\ldots,b_{n}\}$.}
\label{fundgroup}
\end{center}
\end{figure}

The points on an algebraic curve (\ref{fdef}) with 
$f(x,y)=f_{y}(x,y)=0$ can be computed in a standard way as the zeros of the 
discriminant or resultant of $f(x,y)$ and $f_{y}(x,y)$, see 
e.g.~\cite{depa,frkl}  and references therein. 
Their projections to the $x$-sphere $\mathbb{CP}^1$, the base of the covering, are called the \emph{discriminant points}\footnote{An 
algebraic curve of the form (\ref{fdef}) can have singularities, 
i.e., points where in addition to $f(x,y)$ and $f_{y}(x,y)$ also 
$f_{x}(x,y)$ vanishes. Such points as e.g.~double points can have trivial 
monodromies, but are included in the monodromy computation. The 
notion of discriminant points thus includes branch points and 
singular points. Note that an 
algebraic curve has to be desingularized to define a Riemann 
surface, a process not to be discussed here (see for instance 
\cite{depa,frkl} and references therein).} $b_{1},\ldots,b_{n}$  and are assumed  in 
this paper to be given. The task is thus to construct generators $\{\gamma_{k}\}_{k=1}^n$ 
of the fundamental group $\pi_1(\mathbb{CP}^{1}\backslash\{b_{1},\ldots,b_{n}\})$. This means to construct a set of closed contours 
$\gamma_{1},\ldots,\gamma_{n}$,  
all starting at a (finite) common base point $b_{0}$ not being a 
discriminant point, each of the $\gamma_{k}$ encircling exactly one 
discriminant point $b_{k}$ in positive direction and being disjoint from 
other $\gamma_{j}$ everywhere apart from the base point as shown in 
Fig.~\ref{fundgroup}. To take into account a branching of the surface at the point at infinity, a contour $\gamma_{\infty}$ starting and ending at 
$b_{0}$ and encircling all finite discriminant points in negative 
direction is used. We need the contours $\gamma_k$ to satisfy the relation 
\begin{equation}
\gamma_1\gamma_2\dots\gamma_n\gamma_{\infty}={\rm id}.
\label{fundamental}
\end{equation}

In \cite{frkl}, the generators $\{\gamma_k\}$ were constructed in the same way as in the Maple package 
\emph{algcurves}; here we briefly describe this approach. For numerical reasons it is important to stay away 
as much as possible from the discriminant points. Therefore we draw 
small disjoint circles 
centered at the discriminant points, with diameters strictly
smaller than the minimal distance between the discriminant points. Each
circle contains two marked points, the intersections of the circle
with a straight line through the discriminant point parallel to the real
axis. The left and right marked points are denoted by $b_{k}^{(1)}$ 
and $b_{k}^{(2)}$, respectively.  
One of the leftmost marked points is denoted by  $b_0$ and
 chosen to be the base point for 
$\pi_1(\mathbb{CP}^1\setminus\{b_{1},\ldots,b_{n}\})$. Starting from 
this base point straight  lines are drawn to the marked points around 
each of the finite discriminant points. The contours $\gamma_{k}$ 
are formed by these straight lines and the circles around the 
discriminant points. 

\begin{figure}
[!htbp]
\begin{center}
\includegraphics[width=10cm]{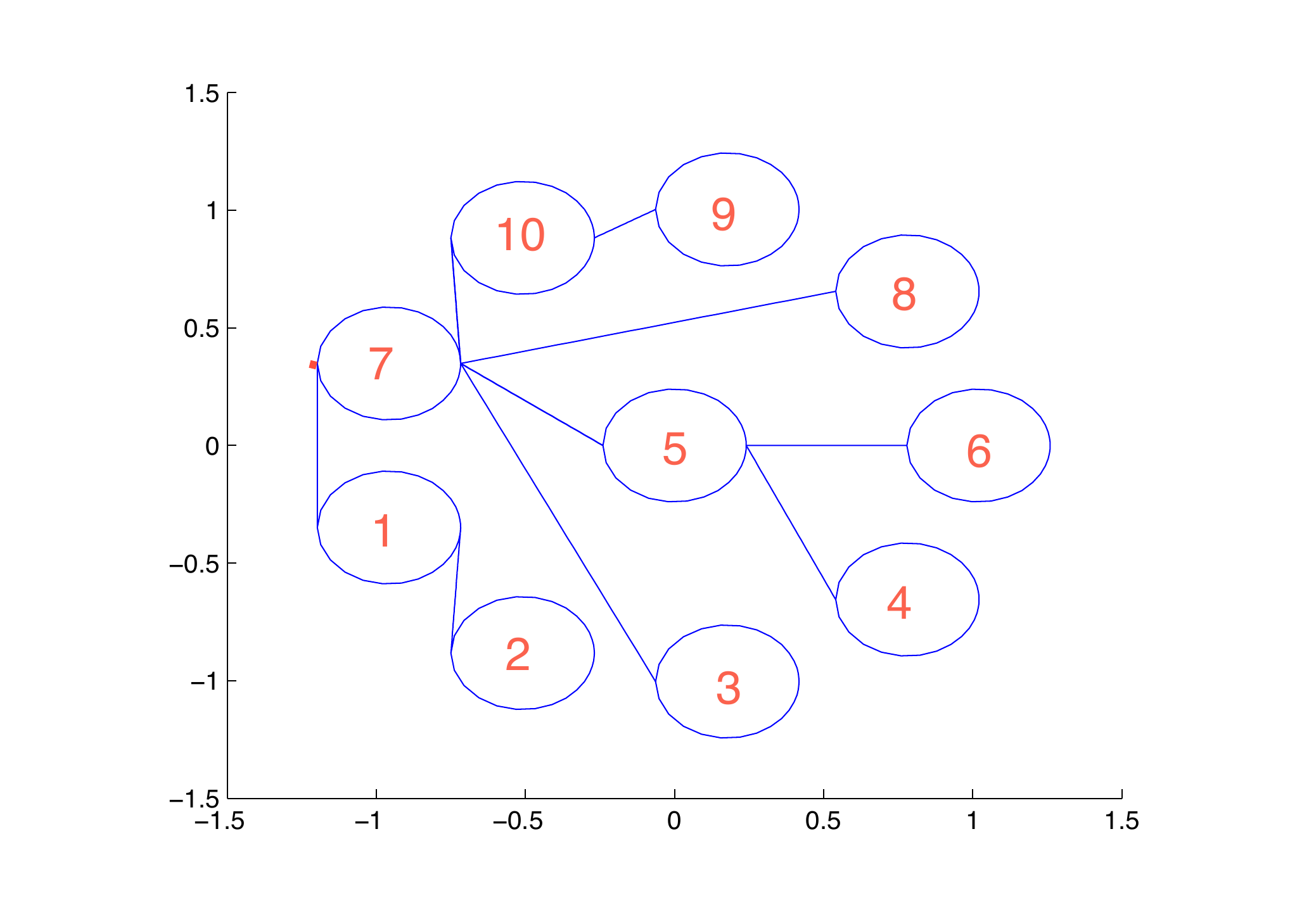} 
\caption{Contours for the monodromy computation for the curve 
$f(x,y)=y^3-2x^3y-x^9=0$ obtained via the deformation approach. The 
base point is marked with a small square.}
\label{treefalt}
\end{center}
\end{figure}

This procedure is best illustrated by an example. Consider the curve 
given by $f(x,y)=y^3-2x^3y-x^9=0$. One quickly 
checks that the discriminant points are given by the roots of 
$x^{9}=2^{5}/3^{3}$ 
and the singular point $x=0$. 
The resulting pattern can be seen in Fig.~\ref{treefalt}.
In this example the contour $\gamma_{7}$ is just the positively 
oriented circle around 
the point $b_{7}$ marked with 7 starting and ending at the base point 
indicated 
with a small square. Similarly the contour $\gamma_{1}$ is formed by 
the line segment from the base point to 
$b_{1}^{(1)}$, the positively oriented full circle around $b_{1}$ and the 
straight line back to $b_{0}$. Drawing a straight line from $b_{0}$ 
to one of the marked points near $b_{10}$ would lead to a line coming 
too close to $b_{7}$. The easiest way to remedy this is to consider 
the straight line between $b_{7}^{(2)}$ and $b_{10}^{(1)}$ instead. 
The contour $\gamma_{10}$ thus consists of the upper half circle  
between $b_{7}^{(1)}$ and $b_{7}^{(2)}$, the line between 
$b_{7}^{(2)}$ and $b_{10}^{(1)}$ and the circle around $b_{10}$. It 
can happen that the distance between a line from $b_{0}$ to a 
point $b_{k}^{(j)}$ and another discriminant point $b_{i}$ is smaller than 
some prescribed minimal distance $\delta$, as would be the case for the line from
$b_{7}^{(2)}$ to $b_{6}^{(1)}$ and $b_{5}$. In this case the contour 
is deformed as follows: instead of this line one considers the line 
between $b_{7}^{(2)}$ and $b_{5}^{(1)}$, the upper half circle around 
$b_{5}$ and the line between $b_{5}^{(2)}$ and $b_{6}^{(1)}$.

Thus if a connecting line comes closer than the distance $\delta$ 
to another discriminant point 
$b_{i}$, it is replaced by lines to and from the $b_{i}^{(j)}$ and a 
half circle around $b_{j}$. Since it is well possible that the new 
lines come also too close to other discriminant points, this procedure has to be 
iterated. A proof that the algorithm terminates has not been given 
(though such a proof should be possible given the finite number of 
problem points). More importantly, the resulting connecting lines will in general not be numerically 
optimal in the sense that they will not have the shortest possible 
lengths as is obvious from Fig.~\ref{treefalt}. 

\begin{figure}
[!htbp]
\begin{center}
\includegraphics[width=10cm]{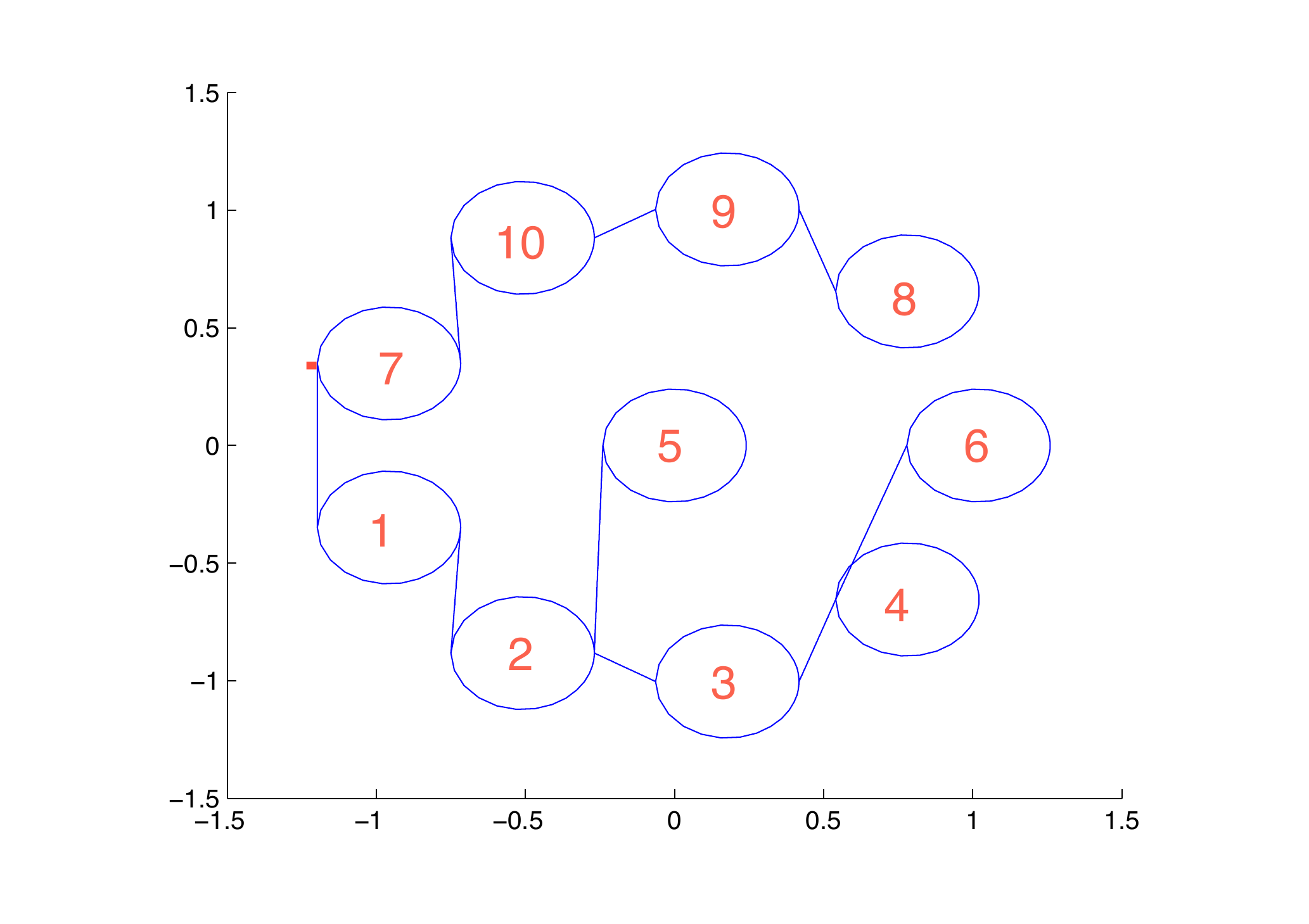} 
\caption{Contours for the monodromy computation for the curve 
$f(x,y)=y^3-2x^3y-x^9=0$ obtained via the minimal spanning tree. The 
base point is marked with a small square.}
\label{treef}
\end{center}
\end{figure}

It is the purpose of this paper to address the outlined problems. Instead of 
deforming the connecting paths, we construct a minimal spanning tree 
having vertices at the discriminant points starting with the $b_{k}$ closest to the 
base point. By construction, edges of this tree will have minimal lengths. The contours $\gamma_k$ are then built as before from 
 line segments between the marked points near discriminant points as they 
appear on the tree and the half circles. The result of this procedure for the same curve as in Fig.~\ref{treefalt} can be seen in Fig.~\ref{treef}.
The tree defines an initial set of contours $\tilde{\gamma}_{k}$, 
$k=1,\ldots,n$ which are numerically 
optimal, but which do not yet satisfy condition 
(\ref{fundamental}).  In a second step, the 
$\tilde{\gamma}_{k}$ will be combined in a way to form a new set of 
contours $\gamma_{k}$, $k=1,\ldots,n$ which satisfy condition 
(\ref{fundamental}).

The paper is organized as follows: In section 2 we describe the 
construction of the minimal spanning tree and the contours $\tilde{\gamma}_{k}$ as well as 
the analytic continuation of the roots $y^{(k)}(b_{0})$, $k=1,\ldots,N$ 
along the contours. In section 3 the found contours are combined in a 
way that they satisfy condition (\ref{fundamental}). In section 4 we compare 
the actual numerical performance of the two approaches for some 
examples.

\section{Contours for integration and minimal spanning tree}
In this section we will explain how to construct the contours which 
 generate the fundamental group of $\mathbb{CP}^{1}$ 
minus the finite discriminant points. These contours will be built from 
a minimal spanning tree, and the roots $y^{(k)}$, $k=1,\ldots,N$ of 
(\ref{fdef}) will be analytically continued along them. The contours will not in 
general satisfy condition (\ref{fundamental}) which will be enforced in 
the next section. 

We assume the finite discriminant points $b_{i}$, $i=1,\ldots,n$ to be 
given. Let $\rho$ be the minimal distance between any two of these 
points, 
\begin{equation*}
    \rho := \min\limits_{
    \begin{smallmatrix}
	i,j=1,\ldots,n\\
	i\neq j
    \end{smallmatrix}}
   (|b_{i}-b_{j}|)\;.
\end{equation*}
For numerical reasons, one has to assume that $\rho$ is 
considerably larger than the rounding error (in Matlab 
with double precision this error is typically of the order $10^{-14}$).
In practice, $\rho$ has to be much larger for the reasons discussed below, see Remark \ref{rmk_resolution}. 
The code issues 
a warning if the ratio of the smallest distance to  the largest 
distance between any two discriminant points is smaller than 
$10^{-4}$, but will typically produce correct results in such 
cases.  The code performs several checks to ensure 
that the obtained results are correct. 

The starting configuration is as follows. Small disjoint circles 
centered at the discriminant points are drawn, with radius $R=\kappa 
\rho$ and $\kappa<1/2$. In Figs.~\ref{treefalt} and \ref{treef} 
we chose $\kappa=1/2.9$ for plotting purposes, for the later computations values of up to 
$\kappa=1/2.1$ are taken\footnote{The value of $\kappa$ is 
essentially  fixed by hand. Since the same number of collocation points is used on 
the half-circles and on the connecting lines between them, the length of the half-circles and the segments of the connecting line between them should ideally be the same to produce a homogeneous numerical resolution over the path. Therefore, a value of 
$\kappa$ close to $1/(\pi+2)$ would be an appropriate choice. However, for an efficient resolution of 
high order singularities, where several sheets of the covering 
coincide, we choose the distance from the path to each critical point to be almost the maximal possible. 
In practice the code uses $\kappa=1/2.1$ to allow for connecting 
lines of positive length between the circles (the Maple package works 
with $\kappa=2/5$).}.  Each
circle contains two marked points, the intersection of the circle
with a straight line through the discriminant point parallel to the real
axis. The marked points on the circle around $b_k$ are
denoted by $b^{(1)}_k$ and $b^{(2)}_k$, where 
${\rm Re}\left\{b_k^{(1)}\right\}<{\rm Re}\left\{b^{(2)}_k\right\}.$
One of the leftmost marked points is chosen to be the base point 
$b_{0}$ for 
$\pi_1(\mathbb{CP}^1\setminus\{b_1,\dots,b_n\})$. The discriminant 
points $\{b_k\}_{k=1}^n$ 
are then ordered according to the ascending complex argument of the
vectors $b_k-b_0$, the argument being measured from $-\pi$ to $\pi$;
if two discriminant points lie on the same ray originating at $b_0$, then
the discriminant point which is closer to the base point is preceding in
the order. This ordering is shown for the studied example in 
Figs.~\ref{treefalt} and \ref{treef} by the numbers in the circles.  

Then a minimal tree originating at the discriminant point $b_{k_{0}}$ closest 
to the base point, $b_{0}=b_{k_{0}}^{(1)}$, is constructed. To this end, all 
distances to the remaining $b_{i}$, $i\neq k_{0}$ are computed, and 
the point 
with the smallest distance, denoted $b_{k_{1}}$, is chosen as the 
next vertex 
on the tree. If there are several points with the smallest distance, 
then the one with the smallest ordering number is chosen. The code 
stores the pair $[k_{0},k_{1}]$ to indicate that the tree starts at 
$b_{k_{0}}$ and that its next vertex is at $b_{k_{1}}$. To obtain the 
next vertex, the distances between both $b_{k_{0}}$ and $b_{k_{1}}$ and the remaining 
$b_{i}$, $i\neq k_{0},k_{1}$ are computed. The smallest distance (in 
case of degeneracies again the point with the smallest ordering 
number is chosen) gives the next vertex $b_{k_{2}}$ of the tree, connected by an edge either to 
$b_{k_{0}}$ or to $b_{k_{1}}$. The code 
stores the pair $[k_{0},k_{2}]$ or $[k_{1},k_{2}] respectively. $Repeating this several 
times one ends up with a tree containing the points $b_{k_{0}}, 
b_{k_{1}},\ldots,b_{k_{m}}$ as vertices. The next vertex on the tree is 
determined as before by computing the minimal distance between points 
already on the tree and points $b_{k}$, $k\neq k_{0},\ldots,k_{m}$ and 
dealing with degeneracies as before. 
Thus by construction, one obtains in this way a minimal spanning tree 
of the $b_{k}$, $k=1,\ldots,n$ originating at $b_{k_{0}}$. The tree 
is not unique because of possible degeneracies of distances between 
the points, but the described algorithm will always produce a 
connected
spanning tree with minimal distances between the points. The result 
of this procedure for the curve in Fig.~\ref{treef} is 
\begin{verbatim}
    paths' =

    7     1     2     3     7    10     9     4     2
    1     2     3     4    10     9     8     6     5,
\end{verbatim}
where it can be seen that there is a connection between 
7 and 1, then between 1 and 2 , between 2 and 3 and so on.  

This tree just indicates in which order the discriminant points 
appear on the paths $\tilde{\gamma}_k$. The actual contours will consist of half 
circles around the $b_{k}$, $k=1,\ldots,n$ and straight lines between 
the points $b_{k}^{(1,2)}$, $k=1,\ldots,2$. Thus in a separate step 
the precise paths will be determined. For each pair of consecutive points 
$b_{k_{0}}$ and $b_{k_{1}}$
appearing on the tree, the connecting lines between 
$b_{k_{0}}^{(1,2)}$ and $b_{k_{1}}^{(1,2)}$ are chosen in a way that 
they intersect the circles around these two points as little as 
possible. To this end the code determines the real part of the 
difference between the two points, $d={\rm Re}(b_{k_{1}}-b_{k_{0}})$. If 
this distance is greater or equal to $R$, the line segment connects
$b_{k_{0}}^{(2)}$ and $b_{k_{1}}^{(1)}$, if it is smaller than $-R$, the 
line will be between $b_{k_{0}}^{(1)}$ and $b_{k_{1}}^{(2)}$, and for 
values between $-R$ and $R$, the points $b_{k_{0}}^{(1)}$ and 
$b_{k_{1}}^{(1)}$ are connected. The result of this procedure is stored in a pair of 
numbers for each pair of neighbouring vertices in the tree. For the example of 
Fig.~\ref{treef} the code gives
\begin{verbatim}
    pathind' =

    1     2     2     2     2     2     2     1     2
    1     1     1     1     1     1     1     1     1.
\end{verbatim}
This has to be read together with the above information given by the variable {\it path}. The two variables together specify the edge $[b_i^{(j)}, b_k^{(l)}]$ of the tree, where the variable {\it path} gives the pair $(i,k)$ and {\it pathind} gives $(j,l)$. 
In the considered example, the first 
line segment is thus between $b_{7}^{(1)}$ and $b_{1}^{(1)}$ as can be seen 
in Fig.~\ref{treef}.
By construction, all connecting lines will have minimal possible 
lengths whilst keeping at least the distance  $\delta=R\sqrt{1-\kappa^{2}}$ 
(see \cite{frkl}) away from  the discriminant points.

The contours $\tilde{\gamma}_{k}$ are then built from these connecting
lines and half circles around discriminant points in the following way.  The
contour $\tilde{\gamma}_{k_1}$ is a contour starting at $b_0$ and
encircling the point $b_{k_1}$ only. Thus the code constructs
$\tilde{\gamma}_{k_1}$ that starts at the base point $b_{0}$, goes
between the points $b_{k}$ in a sequence of connecting lines and half
circles appearing on the tree before the index $k_{1}$, then it
follows positively the circle around $b_{k_{1}}$ and, finally, takes
the same path (minus the circle around $b_{k_{1}}$) back to the base
point. Any discriminant point appearing on this path is bypassed on a
half-circle in positive direction.  The contour $\tilde{\gamma}_{5}$
in Fig.~\ref{treef} starts for instance at the base point
$b_{7}^{(1)}$ with the straight line to $b_{1}^{(1)}$, then the half
circle in positive direction around $b_{1}$ to $b_{1}^{(2)}$, from
there the straight line to $b_{2}^{(1)}$, the half circle around
$b_{2}$ in positive direction to $b_{2}^{(2)}$, from there the
straight line to $b_{5}^{(1)}$, and after a full circle around $b_{5}$
in positive direction the same path from $b_5^{(1)}$ back to the base
point $b_{7}^{(1)}$ in the opposite direction.

\begin{figure}[htb]
  \centering
  \includegraphics[width=0.5\linewidth]{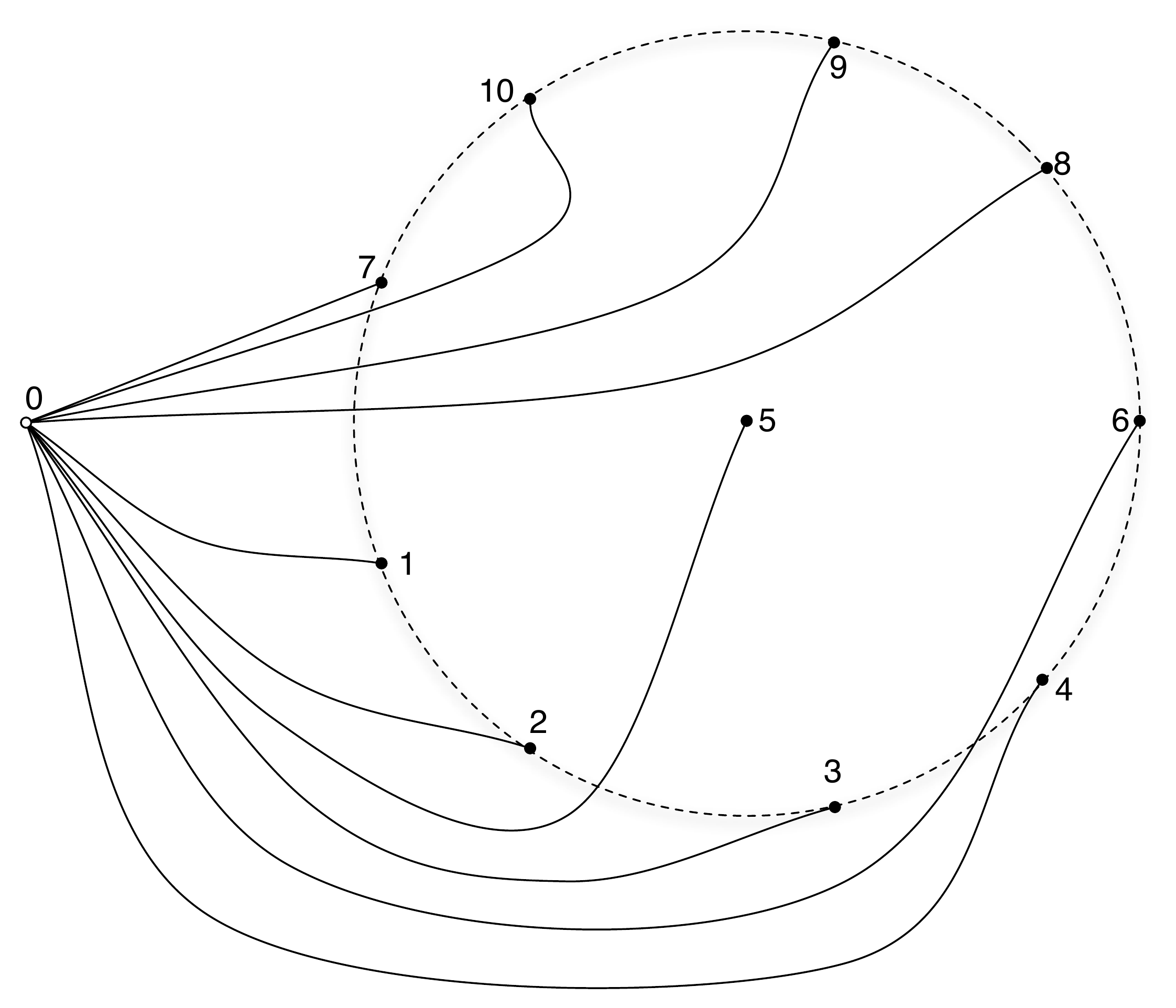}
  \caption{Schematic diagram for the relative position of the 
  contours generated by the minimal tree shown in Fig.~\ref{treef}.}
  \label{fig:contours}
\end{figure}

A schematic view of the relative positions of 
the loops $\tilde\gamma_k$, $k=1,\dots,10$ constructed from the spanning tree in Fig.~\ref{treef} is presented in Fig.~\ref{fig:contours}.

To determine the monodromies for this set of contours, the algebraic
equation (\ref{fdef}) is solved for $y$ at the base point $x=b_{0}$,
which is a generic point of the curve. Thus, there will be $N$
distinct roots $y^{(k)}_{0}$, $k=1,\ldots,N$ at this point which can
be determined numerically with the Matlab function \emph{roots}. The
roots are labeled $1,\ldots,N$, thus numbering the sheets of the
covering of the $x$-sphere $\mathbb{CP}^1$. These roots are then
analytically continued along $\tilde{\gamma}_k$. The analytical
continuation results in the same set of roots $y^{(k)}_{0}$ at $x=b_0$
but in a different order. The permutation of the roots thus obtained
is the monodromy of the Riemann surface along the path
$\tilde{\gamma}_k.$

In order to compute the analytical continuation, we introduce a numerical grid, which amounts to choosing collocation points on each of the 
 line segments and half circles. Since later on the code uses $\tilde{\gamma}_k$ as integration contours to compute integrals of  \emph{holomorphic differentials} (see 
\cite{depa,frkl} how these can be determined)  
to obtain the \emph{periods} of a Riemann surface (integrals of the 
holomorphic differentials along closed contours formed by the 
$\tilde{\gamma}_{k}$), it is convenient 
to choose the collocation points in accordance with the used 
integration scheme. Since we use numerically optimal Gauss-Legendre 
integration, which can be implemented conveniently in Matlab with 
Trefethen's code \cite{trefethen1, trefethenweb}, we take Gauss-Legendre points. 
On each line segment and half circle there will be thus $N_{G}$ points 
(typically $N_{G}$ is between 32 and 128). 

At each of these collocation 
points $x^{c}$, we use \emph{roots} to solve numerically equation (\ref{fdef})  
 to obtain $N$ roots $y^{(k)}(x^{c})$. In general, the 
ordering of these roots will not correspond to the one 
introduced at the base point. Thus the roots are sorted in a way to 
have minimal difference with the roots at the previous collocation 
point, i.e., $|y^{(k)}(x^{c}_{i}) - y^{(k)}(x^{c}_{i-1})| = 
\underset{j=1,\dots,N}{\rm min}|y^{(j)}(x^{c}_{i}) - 
y^{(k)}(x^{c}_{i-1})|$. 
In this way the vector $\vec{y}$ of roots is analytically 
continued along the 
contours $\tilde{\gamma}_{k}$. 

\begin{remark} \rm
    By construction, no discriminant points appear on the contours 
    $\tilde{\gamma}_{k}$ which implies that there will be always $N$ 
    distinct roots $y^{(i)}$ of (\ref{fdef}) on these contours. The roots 
    can be computed in Matlab efficiently with the function 
    \emph{roots} as long 
    as they are well separated. It is known that the computation 
    of almost degenerate zeros of polynomials is an extremely 
    difficult numerical problem, see for instance \cite{zeng} and 
    references therein. In the present context, having almost degenerate 
    roots would mean that the 
    sheets come very close to each other, which is typically the case near high order 
    singularities. Such singularities can be seen as a condensation of 
many double points (the point (0,0) of the example in 
Fig.~\ref{treef} is of this type). If another discriminant point 
comes so close to such a
    singularity that the sheets can no longer be numerically 
    distinguished, i.e., if $\rho$ becomes too small in such a case, the surface cannot be 
    studied with the present code. It is in fact the ability to 
    distinguish the sheets numerically with the \emph{roots} function 
    of Matlab along the contours $\tilde{\gamma}_{k}$ that imposes 
    limitations on
    which curves can be treated by the code.
 Since these limitations depend strongly on the considered curve, 
it is impossible to give a priori limits on the 
applicability of the code.
    \label{rmk_resolution}
\end{remark}

If we start at the base point $b_{0}$ and 
analytically continue the vector of roots $\vec{y}$ with components 
$y^{(i)}$, $i=1,\ldots,N$ as described above along one of the 
contours $\tilde{\gamma}_{k}$, we will in
general obtain a permutation $\sigma_{k}$ of the components of the vector $\vec{y}$
back at the base point,
\begin{equation}
    \sigma_{k}\vec{y}:=(y^{\sigma_{k}(1)}(a),\ldots,y^{\sigma_{k}(N)}(a)).
    \label{perm}
\end{equation}
The group generated by the $\{\sigma_{i}\}_{i=1}^n$  is called the monodromy group
of the covering. The code stores the monodromies $\sigma_i$ in the form of a 
vector of the indices $(\sigma_{i}(1),\ldots,\sigma_{i}(N))$. 
For the curve  in Fig.~\ref{treef} and the set of contours $\tilde{\gamma}_{k}$, 
$k=1,\ldots,n$, one obtains
 the base point
\begin{verbatim}
base =

      -1.2895 + 0.3485i
ybase =

	-0.9546 - 2.8682i
	 1.9591 + 1.1931i
	-1.0044 + 1.6751i
\end{verbatim}
and the monodromies
\begin{verbatim}
    Mon =

	 3     2     1     3     3     1     1     3     1     3
	 2     1     3     2     2     3     3     2     3     2
	 1     3     2     1     1     2     2     1     2     1.
\end{verbatim}
This is to be read in the following way: the vector of roots \emph{ybase} is analytically continued along the loop $\tilde{\gamma}_1$; the result of this continuation is a new vector of roots obtained from \emph{ybase} by permuting the components as specified by the first vector of \emph{Mon}, the permutation $(321)$, i.e.,
starting in the first sheet, one ends up in the third, starting from 
the second one stays there, and starting in the third one ends up in 
the first. 

As was already mentioned, the monodromy at infinity can be computed by 
analytically continuing the vector of roots $\vec{y}$ at the base $b_0$ along a 
closed contour starting and ending at $b_0$ and encircling once 
all finite discriminant points in negative direction as shown in 
Fig.~\ref{fundgroup}. Since the 
radius of such a loop can be very large, a high number of collocation 
points would be needed to obtain the same accuracy as for the loops 
around the finite discriminant points. Thus we will obtain the 
monodromy there from the relation 
$\gamma_{1}\ldots\gamma_{n}=\gamma_{\infty}^{-1}$, see (\ref{fundamental}). 


\section{Generators of the fundamental group}
The set of \emph{initial loops} $\{\tilde{\gamma}_k\}$ constructed in the previous 
section does not satisfy in general condition (\ref{fundamental}), as can be seen from Fig.~\ref{fig:contours} . We call 
the monodromies $\{\tilde{\phi_k}\}_{k=1}^n$ computed along these 
loops the {\it initial monodromies.} 
In this section we  
explain how generators
of the fundamental group of the base of the covering punctured at the
discriminant points $\mathbb{CP}^1\setminus\{b_1,\dots,b_n\}$  
satisfying (\ref{fundamental}) can be 
constructed from the initial loops.


Let us suppose that a permutation $\sigma\in S_n$ exists such that 
\begin{equation}
\label{sigma}
\tilde{\gamma}_{\sigma(1)}\tilde{\gamma}_{\sigma(2)}\dots\tilde{\gamma}_{\sigma(n)}=
\tilde{\gamma}_{\infty}^{-1}.
\end{equation}

Then  there are two  possible ways to proceed. First, the discriminant
points $\{b_k\}$ can be reordered according to $\sigma$, i.e.,
$b_k:=b_{\sigma(k)}$ and the corresponding generators of the
fundamental group are given by $\gamma_k:=\tilde{\gamma}_{\sigma(k)}.$
However, we would like to keep the initial ordering of the branch
points. To this end, the following algorithm is applied to construct
the $\{\gamma_k\}$ satisfying (\ref{fundamental}): (i) if the permutation
$\sigma$ is trivial, then put $\gamma_k=\tilde{\gamma}_k$. Otherwise, 
(ii) let $\sigma(m)$ be the largest number such that $\sigma(m)\neq
m.$ 
Then $\sigma(m+1)<\sigma(m)$ and the loop
$\tilde{\gamma}_{\sigma(m)}$ is redefined as follows:
$\tilde{\gamma}_{\sigma(m)}:= \tilde{\gamma}_{\sigma(m+1)}
\tilde{\gamma}_{\sigma(m)}\tilde{\gamma}_{\sigma(m+1)}^{-1}$ (the loop $\tilde{\gamma}_{\sigma(m+1)}^{-1}$ is traced first, then $\tilde{\gamma}_{\sigma(m)}$ followed by $\tilde{\gamma}_{\sigma(m+1)}$, see also Fig.~\ref{fig:rearrange} below). The
permutation $\sigma$ is then composed with the transposition swapping
$\sigma(m)$ and $ \sigma(m+1)$, i.e.,  $\sigma:=(\sigma(m)
\sigma(m+1))\circ \sigma$. After this, the algorithm is reiterated. 

Let us now find the permutation $\sigma$ from (\ref{sigma}) imposed
by the minimal tree and the construction of the initial loops
$\{\tilde{\gamma}_k\}$.

 Consider two loops $\gamma$, $\delta \in
\pi_1(\mathbb{CP}^1\setminus\{b_1,\dots,b_n\}, b_0),$ more precisely, 
consider the parts from $b_0$ to the respective points they encircle. 
These might have a common part close to the base point like some of
the initial loops $\{\tilde{\gamma}_k\}$ do. At the point when
$\gamma$ and $\delta$ separate, denote by $\overrightarrow{\gamma}$
and $\overrightarrow{\delta}$ the tangent vectors to the loops. Then
the orientation of the pair ($\overrightarrow{\gamma}$,
$\overrightarrow{\delta}$) indicates the relative position of the two
loops. 

\begin{lemma}
\label{lm_loops}
Suppose now that a set of generators $\{\gamma_k\}_{k=1}^n$ of the
group  $\pi_1(\mathbb{CP}^1\setminus\{b_1,\dots,b_n\},
b_0)$ is such that (i) each loop encircles only one puncture; (ii)
the pair ($\overrightarrow{\gamma}_k$,
$\overrightarrow{\gamma}_{k+1}$) is positively oriented at any point
of separation of ${\gamma}_k$ and ${\gamma}_{k+1}$ for all
$k=1,\dots,n-1;$ (iii)  
the loops $\gamma_k$ do not intersect each other apart from the base
point (there exist representatives in the corresponding homotopy
classes which do not intersect). Then the loops $\{\gamma_k\}$
satisfy (\ref{fundamental}). 
\end{lemma}
The proof of this lemma is obvious.

By construction, the initial loops $\{\tilde{\gamma}_k\}$ satisfy
conditions (i) and (iii) of the lemma. Therefore, we are looking for
the permutation $\sigma\in S_n$ such that the pairs
($\overrightarrow{\gamma}_{\sigma(k)}$,
$\overrightarrow{\gamma}_{\sigma(k+1)}$) be positively oriented at
the corresponding points of  separation of loops for all
$k=1,\dots,n-1$. 

Let us introduce some notation.   
Whenever the path contains a sequence of edges of the type 
$[\dots,b^{(1)}_j] \; [b^{(1)}_j,\dots]$ or $[\dots,b^{(2)}_j] \; 
[b^{(2)}_j,\dots]$, the point $b_j^{(1)}$ or $b_j^{(2)}$,  is called a {\it 
v-point}. 
 The only such point in Fig.~\ref{treef} is $b_{4}^{(1)}$. 
We call  a part of a single branch of
a tree without v-points a {\it string}. A {\it node} is a vertex
where several branches meet. In what follows, the v-points are
considered as nodes, where the corresponding discriminant point becomes a
separate branch consisting of only one point. We call a node {\it
simple} if all of its descendants are strings. 
 
 In order to find an algorithm which produces the permutation $\sigma$ from (\ref{sigma})
for the given minimal tree, we first discuss two particular types of
trees. 

\begin{paragraph}{ \bf I} In the case when the minimal tree is a string,
the permutation $\sigma=(\sigma(1)\dots\sigma(n))$ is given by the
sequence of labels of the branch points read from the end of the
string towards the base point $b_0$ so that  $b_0 =
b^{(1)}_{\sigma(n)}$. This follows directly by construction of the
initial loops.  
\end{paragraph}

\begin{paragraph}{\bf II}
Suppose the tree contains only one node which coincides with the base
point $b_0.$ Suppose there are $m$ branches meeting at the node, each
of which is a string. To each branch of the tree the algorithm
associates a sequence $s^{i}=(s^i_1,\dots,s^i_{n_i})$, $i=1,\dots,m$
of numbers like above - a sequence of labels of the discriminant points
read from the end of the string towards the node. This sequence
indicates the order in which the loops  $\{\tilde{\gamma}_k\}$ should be
composed to give a positively oriented loop around all points contained in the string.
Now we need to decide on the relative position of the branches at the
node, i.e., a relative position of the vectors
$\overrightarrow{\tilde{\gamma}}_{s^i_{1}}$, $i=1,\dots,m$ at
$b_0$. We order the vectors according to the ascending angle they
make with the horizontal ray going from the base point $b_0$ to the
left, the angle is measured from $0$ to $2\pi.$ This order is
expressed as a permutation $\rho\in S_m$ of the indices attached to
the branches, i.e., the sequences $s^i$ are ordered as follows:
$s^{\rho(1)}, \dots, s^{\rho(m)}$. Now the required permutation
$\sigma$ is obtained by writing the sequences $s^i$ one after the
other in the order they appear at the node:
$\sigma=(\sigma(1)\dots\sigma(n)):=(s^{\rho(1)}_1\dots
s^{\rho(1)}_{n_{\rho(1)}} s^{\rho(2)}_1\dots s^{\rho(2)}_{n_{\rho(2)}} \dots
s^{\rho(m)}_1 \dots s^{\rho(m)}_{n_{\rho(m)}}).$
\end{paragraph}

%
Now we are in a position to present the complete algorithm.

\begin{paragraph}{\textbf{Algorithm}.} In the general case, the algorithm
first identifies the end points of the tree (those without
descendants) and the nodes (including the v-points). Then starting at
each of the end points, the algorithm forms a sequence of labels of
the discriminant points going from the current point to its parent until it
hits a node.  This process results in a set of all simple nodes and a
set of sequences of numbers attached to every such node. 
 For each of the simple nodes, the procedure from case {\bf II} is
performed where, if the node is different from $b_0,$ instead of the
horizontal ray going left from the base point,  the line of arrival
to the node with the reversed orientation is taken (the line of arrival is the line segment from the previous marked point on the path to the current one). The result of
this procedure is a sequence $\bf{s}$ of numbers at each node which
indicates the order in which the loops $\{\tilde{\gamma}_k\}$ should
be composed to obtain a positively oriented loop encircling all descendants of the given
simple node and only them. Thus the part of the tree descending from
a simple node can be treated as a string in which the points are
arranged in the obtained order $\bf{s}$. Therefore, after the
ordering of the loops at all simple nodes is done, each simple node
is considered as an end point of the tree whose label is given by the
sequence of numbers $\bf{s}$. Then the algorithm is reiterated. 
\end{paragraph}
 
We illustrate the algorithm on the example of Fig.~\ref{treef}. The 
code first identifies the endpoints 
\begin{verbatim}
endpoints =

	 8
	 6
	 5,
\end{verbatim}
and the nodes 
\begin{verbatim}
nodes =

	 7     2
	 1     2,
\end{verbatim}
corresponding to the points $b_{7}^{(1)}$ (the base point) and 
$b_{2}^{(2)}$. For technical reasons the v-points are not identified 
at this stage. Starting from the endpoints, the code then traces the 
branches until it hits the first node on each branch. At each point, 
it is checked whether the point is a v-point. If such 
a point is reached, the standard order procedure at a node is 
applied. Each point not being a v-point on such a 
branch is listed in the order of appearance.
Thus at the v-point $b_4^{(1)}$, the code considers two branches $6$ 
and $4$ and places $4$ in front of $6$ since the angle between the 
reversed arrival line $\overline{b_4^{(1)}b_3^{(2)}}$ and 
$\overline{b_4^{(1)}b_4}$ is smaller than the one between $\overline{b_4^{(1)}b_3^{(2)}}$ and $\overline{b_4^{(1)}b_6^{(1)}}$. At the 
node $b_{2}^{(2)}$ we thus get the two strings
\begin{verbatim}
tree{2} =

         4     6     3

tree{3} =

         5.
\end{verbatim}
At the node $b_2^{(2)}$, these strings are combined into a single string, where the sequence \emph{tree\{{\rm2}\}} comes in front of \emph{tree\{\rm 3\}} because the angle between the reversed arrival line $\overline{b^{(2)}_{2}b_{2}^{(1)}}$ and $\overline{b^{(2)}_{2}b_{3}^{(1)}}$ is smaller than that between $\overline{b^{(2)}_{2}b_{2}^{(1)}}$ and $\overline{b^{(2)}_{2}b_{5}^{(1)}}$.
%
This leads 
to 
\begin{verbatim}
Tree =

         4     6     3     5         
\end{verbatim}
This branch and the one from the remaining endpoint are then traced 
to the base point, where we get 
\begin{verbatim}
tree{1} =

        8     9    10     7
	
tree{2} =

        4     6     3     5     2     1.	
\end{verbatim}
Again the two strings are combined at the base point according to the 
angles of the connecting lines to 
give
\begin{verbatim}
Tree =

        4     6     3     5     2     1     8     9    10     7.
\end{verbatim}
In words, this string gives the relative `position' of homotopy 
classes of the loops which can also be seen from Fig.~\ref{fig:contours}: 
in this sense the contour
$\tilde{\gamma}_{10}$ is entirely to the `left' of 
$\tilde{\gamma}_{k}$, $k\neq 7,10$ as wanted, but it is to the `right' of 
$\tilde{\gamma}_{7}$. Thus the contour $\gamma_{10}$ 
will be obtained by  conjugating $\tilde{\gamma}_{10}$  with $\tilde{\gamma}_7$ to get 
$\gamma_{10}:=\tilde{\gamma}_{7}\tilde{\gamma}_{10}\tilde{\gamma}_{7}^{-1}$ (by this notation we mean that the contour $\tilde{\gamma}_{7}^{-1}$ is traced first, then $\tilde{\gamma}_{10}$ followed by $\tilde{\gamma}_{7}$). 
The action of such a conjugation is illustrated in 
Fig.~\ref{fig:rearrange}: the situation before the conjugation 
can be seen on the right, and the effect of the conjugation on the 
left.
\begin{figure}[htb]
  \centering
  \includegraphics[width=0.9\linewidth]{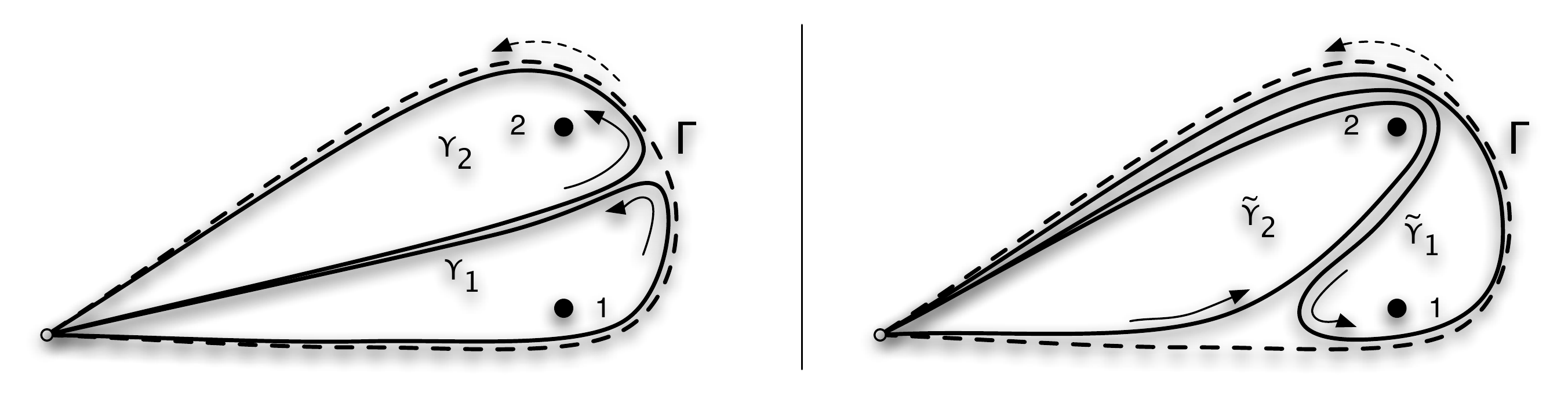}
  \caption{The contour $\Gamma$ surrounding the two points $1$ and $2$ is represented in two different ways by contours around the individual points. On the left, $\Gamma = \gamma_2 \gamma_1$ while $\Gamma = \tilde\gamma_1 \tilde\gamma_2$ on the right. Since $\gamma_2=\tilde\gamma_2$ it follows that $\tilde\gamma_1= \gamma_2\gamma_1 \gamma_2^{-1}$.  }
  \label{fig:rearrange}
\end{figure}

The code 
identifies the largest number $m$ in the string \emph{Tree}. If it is 
in the right-most position, this number is deleted from $Tree$. If there is a number $k$ to the right, 
the loop $\tilde{\gamma}_{m}$ and the monodromy 
$\tilde{\phi}_{m}$ are redefined by 
$\tilde{\gamma}_{m}:=\tilde{\gamma}_{k}\tilde{\gamma}_{m}\tilde{\gamma}_{k}^{-1}$ and  
$\tilde{\phi}_{m}:=\tilde{\phi}_{k}\tilde{\phi}_{m}\tilde{\phi}_{k}^{-1}$. 
The numbers $m$ and $k$ are then swapped in $Tree$. This procedure 
is repeated until $m$ is in the right-most position. It 
is then deleted from $Tree$ and the procedure is repeated until there 
is only one element in $Tree$ left. The resulting loops and 
monodromies are the desired $\gamma_{k}$ and $\phi_{k}$. 
Finally the monodromy at infinity is obtained from 
relation (\ref{fundamental}). For the considered example this leads 
to 
\begin{verbatim}
    Mon =

	 1     2     2     3     2     1     3     3     1     3
	 3     1     1     2     1     3     2     2     3     2
	 2     3     3     1     3     2     1     1     2     1.    
\end{verbatim}
In this example infinity is a singular point with trivial monodromy 
which is why the code gives no monodromy at infinity (it would appear at 
position $n+1$). 
We note that it is possible to compute the genus $g$, 
the only topological invariant of a Riemann surface, from the 
monodromies via the Riemann-Hurwitz formula,
\begin{equation*}
    g = 1-N + \frac{1}{2}\sum_{i=1}^{N_{B}}\beta_{i}
\;,
\end{equation*}
where $N$ is the total number of sheets, $\beta_{i}$ is the branching number, the number of sheets connected at a point on the covering minus 1, and where $N_{B}$ is number of 
discriminant points on the covering. For the studied example one finds 
thus $g=3$.

\section{Performance of the code}

We have described two algorithms for computing monodromies. The first constructs contours $\gamma_k$ by the deformation approach, while the second achieves this from a spanning tree construction. In order to judge the performance of both approaches we
will compute characteristic quantities of a Riemann surface for several examples.

It is known (see for instance the standard literature on Riemann 
surfaces such as \cite{springer}) that the space of holomorphic one-forms 
$\omega$ of a surface of genus $g$ is $g$-dimensional. For the homology of the 
surface one can introduce a canonical basis of cycles $a_{i}$, 
$b_{i}$, $i=1,\ldots,g$ such that $a_{i}\circ b_{j}=\delta_{ij}$. 
These $a$- and $b$-cycles can be obtained from the loops 
$\gamma_{k}$, $k=1,\ldots,n$ via an algorithm by Tretkoff and 
Tretkoff \cite{tret}.
For normalized holomorphic one-forms such that
$\oint_{a_{i}}^{}\omega_{j}=\delta_{ij}$, the matrix of $b$-periods 
$\mathbb{B}_{ij}=\oint_{b_{i}}^{}\omega_{j}$ is a Riemann matrix, 
it has a positive definite imaginary part and is symmetric. 
Thus for a given basis of the holomorphic one-forms, the code computes 
the periods from the integrals along the $\tilde{\gamma}_{k}$, 
$k=1,\ldots,n$ via Gauss-Legendre integration. The found numerical 
Riemann matrix will not be exactly symmetric. Since the symmetry of 
$\mathbb{B}$ is not imposed, its asymmetry is a strong test of the 
quality of the numerics. In the following we will use the norm 
$\Delta$ of 
$\mathbb{B}-\mathbb{B}^{T}$ (the eigenvalue  of the 
matrix having the largest absolute value)
as a measure of the numerical error. 
We take two codes which are identical except for the part 
where the contours $\gamma_k$ are generated and compute their performance for typical examples.

As already stated, the curve of Fig.~\ref{treefalt} and \ref{treef} 
has genus 3. A basis of the holomorphic one-forms is given by 
$x^{3}/f_{y}(x,y)$, $x^{4}/f_{y}(x,y)$, and $xy/f_{y}(x,y)$. 
The errors we obtain  for $\kappa=1/2.9$ are given in Table \ref{table1}.
\begin{table}[hbp]
    \centering
    \begin{tabular}{|c|c|c|}
	\hline
	$N_{G}$ & $\Delta_{def}$ & $ \Delta_{st}$\\
	\hline
	8 & $2.14*10^{-5}$ & $1.13*10^{-4}$  \\
	\hline
	16 & $8.15*10^{-9}$ & $1.55*10^{-9}$  \\
	\hline
	32& $1.61*10^{-13}$ & $1.63*10^{-15}$  \\
	\hline
    \end{tabular}
    \vspace{3mm}
    
    \caption{Norm of $\mathbb{B}-\mathbb{B}^{T}$ for the curve 
$f(x,y)=y^3-2x^3y-x^9=0$ for the deformation 
    approach ($\Delta_{def}$) and the spanning tree ($\Delta_{st}$) 
    in dependence of the number $N_G$ of collocation points on each segment 
    of the $\gamma_{k}$.}
    \label{table1}
\end{table}

It can be seen that the error shows the expected spectral convergence 
in both cases, but that the spanning tree gives a numerical error 
almost two orders of magnitude better than the deformation approach 
except for $N_{G}=8$ where the resolution is too low in both cases. 
It is remarkable that machine precision can be reached with this 
method with just $32$ collocation points on each segment of the 
contours. The whole computation takes just 0.5s on a laptop in this 
case.

\begin{figure}
[!htbp]
\begin{center}
 \includegraphics[width=6cm]{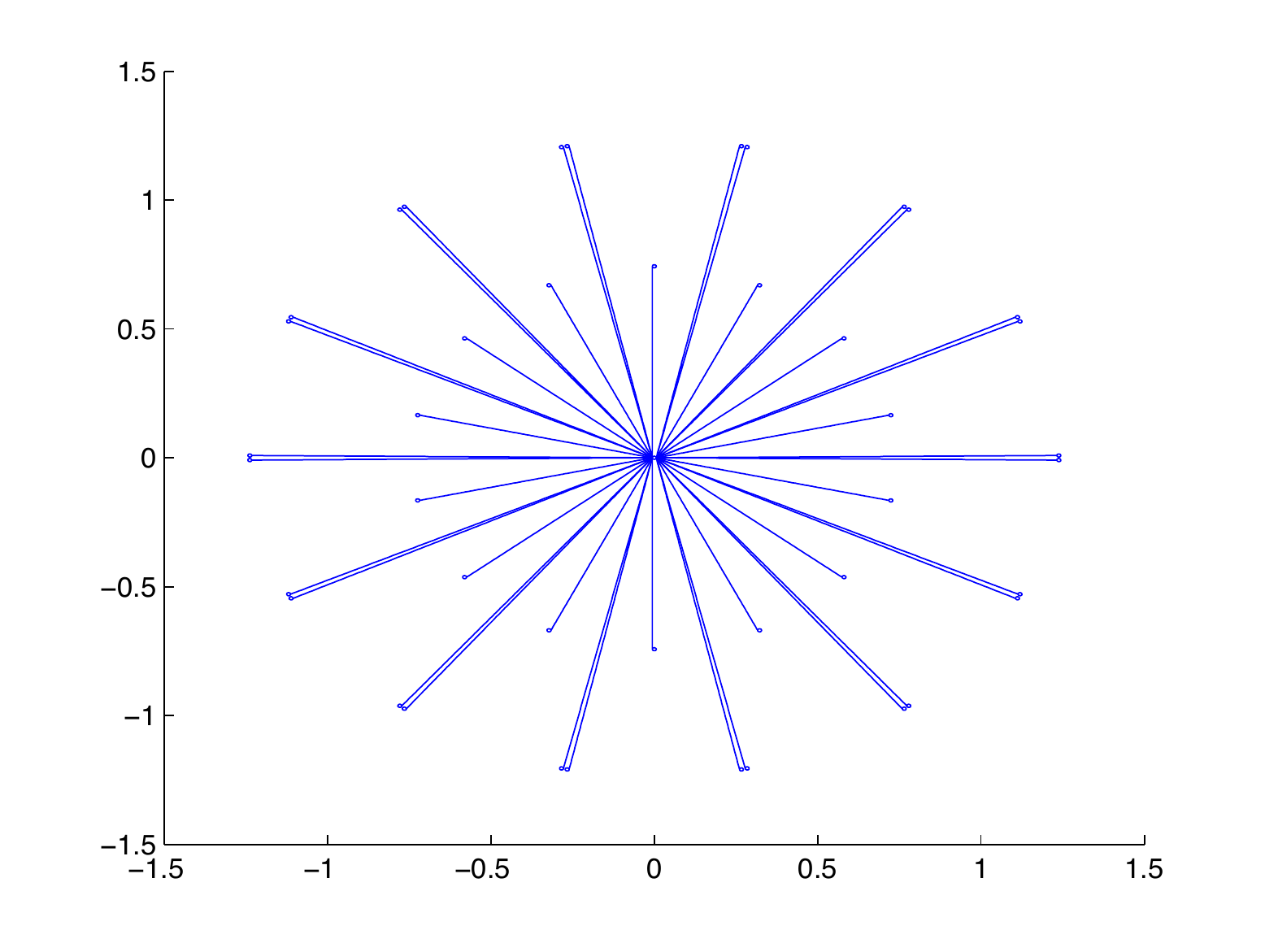}
 \includegraphics[width=6cm]{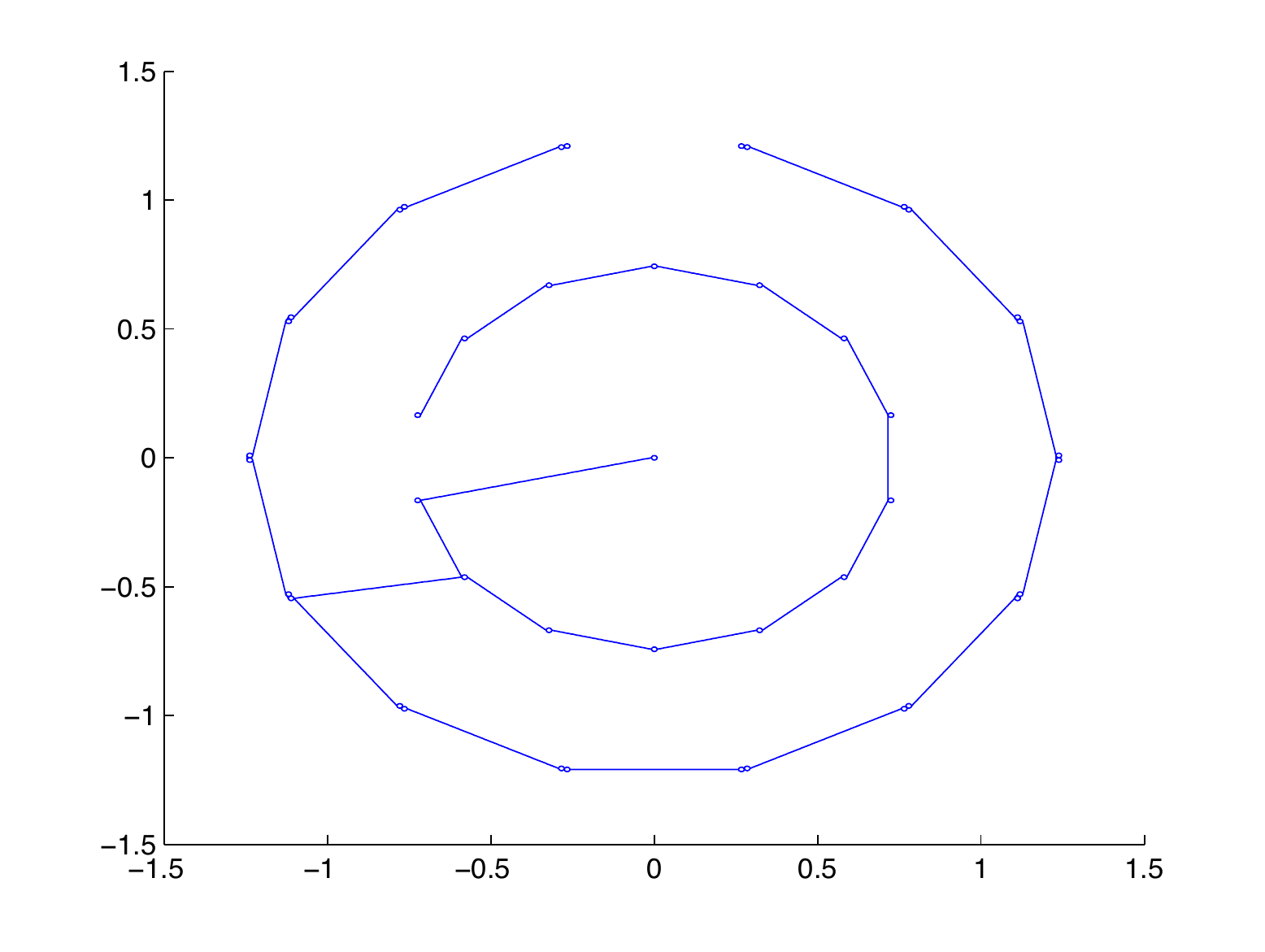} 
\caption{Loops for the monodromy computation for the curve 
$f(x,y):=y^9+2x^2y^6+2x^4y^3+x^6+y^2=0$ obtained with a deformation 
approach with base point close to $x=0$ on the left and with a 
spanning tree on the right.}
\label{treebig}
\end{center}
\end{figure}

The advantage of the spanning tree is more visible for more 
involved curves such as $f(x,y):=y^9+2x^2y^6+2x^4y^3+x^6+y^2=0$. This 
curve of genus 16 has 43 finite discriminant points with  minimal 
distance $\rho=0.018$ 
between them. The monodromy computation is extremely demanding in 
this case. The  points in the outer 
ring in Fig.~\ref{treebig} represent pairs of discriminant points of 
the curve separated by a distance of only $0.018$. For the 
deformation approach we have chosen the base point close to the 
geometric center of the discriminant points, i.e., close to the point 
$x=0$. This gives shorter integration paths and was used in general 
for the deformation approach in \cite{frkl}. For the spanning tree 
the choice of the base point has no influence on the length of the 
connecting lines since we use a minimal tree. 

The code produces the values of the norm of the 
antisymmetric part of the computed Riemann matrices given in Table 
\ref{table2}. The 
computation with $N_{G}=64$ takes $20s$ in the latter case.
\begin{table}[hbp]
    \centering
    \begin{tabular}{|c|c|c|}
	\hline
	$N_{G}$ & $\Delta_{def}$ & $ \Delta_{st}$\\
	\hline
	32 & $\ldots$ & $5.49*10^{-6}$  \\
	\hline
	64 & $6.24*10^{-7}$ & $9.37*10^{-12}$  \\
	\hline
	128 & $3.66*10^{-11}$ & $1.28*10^{-13}$  \\
	\hline
    \end{tabular}
    \vspace{3mm}
    
    \caption{Norm of $\mathbb{B}-\mathbb{B}^{T}$  for the curve 
$f(x,y):=y^9+2x^2y^6+2x^4y^3+x^6+y^2=0$ for the deformation 
    approach ($\Delta_{def}$) and the spanning tree ($\Delta_{st}$) 
    in dependence of the number of collocation points on each segment 
    of the $\gamma_{k}$.}
    \label{table2}
\end{table}
The deformation approach did not produce a result for $N_{G}=32$. 
More importantly the spanning tree needs in this case just half the 
number of modes to reach the same precision as the deformation 
approach until both reach the saturation level. This implies that 
a factor 2 in allocated resources and CPU time can be gained with this approach 
which allows consequently the study of more involved curves. 

%



\begin{thebibliography}{99}

\bibitem{BK} Bobenko, A.I.: Introduction to Compact Riemann Surfaces, 
In Bobenko, A.I., and Klein, C. (ed.), `Computational Approach to Riemann 
Surfaces', Lect. Notes Math. \textbf{2013} (2011).


    \bibitem{algebro}Belokolos, E.D., Bobenko, A.I.,  Enolskii, V.Z.,
    Its, A.R., 
      Matveev, V.B.: Algebro-geometric approach to nonlinear integrable 
      equations. Springer, Berlin (1994)

\bibitem{bo} Bobenko, A., Bordag, L.: 
Periodic multiphase solutions to the Kadomtsev-Petviashvili equation.
J. Phys. A: Math. Gen., \textbf{22}, 1259 (1989)
     

\bibitem{couv}Couveignes, J.-M.: Tools for the computation of 
families of coverings, In Aspects of Galois theory (Gainesville, FL, 1996), 
38-65, London Math. Soc. Lecture Note Ser., 256, (Cambridge Univ. 
Press, Cambridge, 1999).

\bibitem{deho}
Deconinck, B., {v}an Hoeij, M.:
\newblock Computing {R}iemann matrices of algebraic curves.
\newblock Physica {D}, \textbf{152--153}, 28--46 (2001)

\bibitem{depa}Deconinck, B.~ and Patterson, M.: Computing with plane 
algebraic curves,In Bobenko, A.I., and Klein, C. (ed.), `Computational Approach to Riemann 
Surfaces', Lect. Notes Math. \textbf{2013} (2011).

\bibitem{cam}Frauendiener, J.,  Klein, C.: Hyperelliptic
  theta-functions and spectral methods. J. Comp. Appl. Math.,
  \textbf{167}, 193 (2004)

\bibitem{lmp}Frauendiener, J.,  Klein, C.: Hyperelliptic
  theta-functions and spectral methods: KdV and KP solutions,
  Lett. Math. Phys., \textbf{76}, 249-267 (2006)
		
\bibitem{frkl}Frauendiener, J.,  Klein, C.: Algebraic curves and 
Riemann Surfaces in Matlab, In Bobenko, A.I., and Klein, C. (ed.), `Computational Approach to Riemann 
Surfaces', Lect. Notes Math. \textbf{2013} (2011).



\bibitem{cpam}Grava, T., Klein, C.: Numerical solution of the small
  dispersion limit of Korteweg de Vries and Whitham equations,
  Comm. Pure Appl. Math., \textbf{60}, 1623-1664 (2007)

\bibitem{klkoko}Klein, C., Kokotov, A.,  Korotkin, D.: Extremal
  properties of the determinant of the Laplacian in the Bergman metric
  on the moduli space of genus two Riemann surfaces.  
  Math. Zeitschr. \textbf{261}(1), 73--108 (2009)



\bibitem{ernstbook}Klein, C.,   Richter, O.: Ernst Equation and
  Riemann Surfaces. Lecture Notes in Physics \textbf{685}, Springer, 
  Berlin  (2005)


\bibitem{lale}Lax, P.D.,  Levermore, C.D.: The small dispersion
    limit of the Korteweg de Vries equation, I,II,III. Comm. Pure
    Appl. Math., {\bf 36}, 253-290, 571-593, 809-830  (1983)


\bibitem{Sarnak}Quine, J.R., Sarnak, P. (ed): Extremal Riemann
  surfaces. Contemporary Mathematics, {\bf 201} AMS (1997)

\bibitem{schmies}Schmies, M.: Computing Poincar\'e Theta Series for 
Schottky Groups, In Bobenko, A.I., and Klein, C. (ed.), `Computational Approach to Riemann 
Surfaces', Lect. Notes Math. \textbf{2013} (2011).

\bibitem{springer}
Springer, G.:
\newblock Introduction to {R}iemann surfaces.
\newblock Addison-Wesley Publishing Company, Inc., Reading, Mass. 
(1957)

\bibitem{trefethen1}Trefethen, L.N.: Spectral Methods in
    Matlab. SIAM, Philadelphia, PA (2000)

\bibitem{tret}Tretkoff, C.L., Tretkoff, M.D.: Combinatorial
	group theory, Riemann surfaces and differential equations.
      Contemporary Mathematics, \textbf{33}, 467--517 (1984)

\bibitem{trefethenweb} www.comlab.ox.ac.uk/oucl/work/nick.trefethen

\bibitem{zeng}Zeng, Z.: Computing multiple roots of inexact
  polynomials. Math. Comp. \textbf{74}, 869-903 (2004)

\end{thebibliography}
\end{document}